\documentclass[twocolumn]{aastex631}
\pdfoutput=1 % for arXiv submission
\usepackage{amsmath}
\usepackage{apjfonts}
\usepackage{hyperref}

\newcommand{\msunit}{m\,s$^{-1}$}
\newcommand{\ms}{\,\msunit}

\newcommand{\HarvardPhysics}{Department of Physics, Harvard University, 17 Oxford Street, Cambridge MA 02138, USA}

\newcommand{\CfA}{Center for Astrophysics | Harvard \& Smithsonian, 60 Garden Street, Cambridge, MA 02138, USA}

\newcommand{\NASASaganFellow}{NASA Sagan Fellow}

\newcommand{\CHEOPSFellow}{CHEOPS Fellow, SNSF NCCR-PlanetS}

\newcommand{\UKRIFellow}{UKRI Future Leaders Fellow}

\newcommand{\GenevaObservatory}{Observatoire de Gen\`eve, Universit\'e de Gen\`eve, 51 chemin des Maillettes, 1290 Versoix, Switzerland}

\newcommand{\StAndrewsPhysicsAstronomy}{Centre for Exoplanet Science, SUPA, School of Physics and Astronomy, University of St Andrews, St Andrews KY16 9SS, UK}

\newcommand{\BelfastMathPhysics}{Astrophysics Research Centre, School of Mathematics and Physics, Queen's University Belfast, BT7 1NN, Belfast, UK}

\newcommand{\PadovaPhysicsAstronomy}{Dipartimento di Fisica e Astronomia ``Galileo Galilei,'', Universit{\`a} di Padova, Vicolo dell`Osservatorio 3, I-35122 Padova, Italy}

\newcommand{\CavendishLab}{Astrophysics Group, Cavendish Laboratory, J.J. Thomson Avenue, Cambridge CB3 0HE, UK}

\newcommand{\EdinburghAstronomy}{SUPA, Institute for Astronomy, Royal Observatory, University of Edinburgh, Blackford Hill, Edinburgh EH93HJ, UK}

\newcommand{\EdinburghExoplanetCenter}{Centre for Exoplanet Science, University of Edinburgh, Edinburgh, UK}

\newcommand{\INAFTorino}{INAF-Osservatorio Astrofisico di Torino, via Osservatorio 20, I-10025 Pino Torinese, Italy}

\newcommand{\INAFPalermo}{INAF-Osservatorio Astronomico di Palermo, Piazza del Parlamento 1, I-90134 Palermo, Italy}

\newcommand{\INAFBrenaBaja}{INAF-Fundacion Galileo Galilei, Rambla Jose Ana Fernandez Perez 7, E-38712 Brena Baja, Spain}

\newcommand{\INAFCagliari}{INAF-Osservatorio Astronomico di Cagliari, Via della Scienza 5, I-09047 Selargius CA, Italy}

\newcommand{\INAFBrera}{INAF-Osservatorio Astronomico di Brera, Via E. Bianchi 46, I-23807 Merate (LC), Italy}

\newcommand{\MarylandPhysics}{Department of Physics, University of Maryland, College Park, MD 20742, USA}

\newcommand{\MarylandECE}{Department of Electrical and Computer Engineering, University of Maryland, College Park, MD 20742, USA}

\newcommand{\MarylandQTC}{Quantum Technology Center, University of Maryland, College Park, MD 20742, USA}

\newcommand{\KavliInstitute}{Kavli Institute for Cosmology, University of Cambridge, Madingley Road, Cambridge CB3 0HA, UK}

\newcommand{\Exeter}{Astrophysics Group, University of Exeter, Exeter EX4 2QL, UK}

\newcommand{\WarwickPhysics}{Department of Physics, University of Warwick, Coventry, CV4 7AL, UK}

\shorttitle{Exoplanet Detection Limits Using GPs and the HARPS-N Solar RVs}
\shortauthors{N. Langellier et al.}

\begin{document}

%%% TITLE %%%
\title{Detection Limits of Low-mass, Long-period Exoplanets Using Gaussian Processes Applied to HARPS-N Solar RVs}

%%% AUTHORS %%%
% TIER ONE

\author[0000-0003-2107-3308]{N. Langellier}
\altaffiliation[Corresponding author: ]{\href{mailto:nlangellier@gmail.com}{nlangellier@gmail.com}}
\affiliation{\HarvardPhysics}
\affiliation{\CfA}

\author[0000-0001-5446-7712]{T. W. Milbourne}
\affiliation{\HarvardPhysics}
\affiliation{\CfA}

\author[0000-0001-5132-1339]{D. F. Phillips}
\affiliation{\CfA}

\author[0000-0001-9140-3574]{R. D. Haywood}
\altaffiliation{\NASASaganFellow}
\affiliation{\CfA}
\affiliation{\Exeter}

\author[0000-0001-7032-8480]{S. H. Saar}
\affiliation{\CfA}

\author[0000-0001-7254-4363]{A. Mortier}
\affiliation{\CavendishLab}
\affiliation{\KavliInstitute}

\author[0000-0002-6492-2085]{L. Malavolta}
\affiliation{\PadovaPhysicsAstronomy}

\author[0000-0002-8039-194X]{S. Thompson}
\affiliation{\CavendishLab}

\author[0000-0002-8863-7828]{A. Collier Cameron}
\affiliation{\StAndrewsPhysicsAstronomy}

\author[0000-0002-9332-2011]{X. Dumusque}
\affiliation{\GenevaObservatory}

%% TIER TWO

\author[0000-0001-8934-7315]{H. M. Cegla}
\altaffiliation{\CHEOPSFellow}
\altaffiliation{\UKRIFellow}
\affiliation{\GenevaObservatory}
\affiliation{\WarwickPhysics}

%\author{J. Costes}
%\affiliation{\BelfastMathPhysics}

\author[0000-0001-9911-7388]{D. W. Latham}
\affiliation{\CfA}

\author[0000-0002-2218-5689]{J. Maldonado}
\affiliation{\INAFPalermo}

\author{C. A. Watson}
\affiliation{\BelfastMathPhysics}

%TIER THREE

\author[0000-0002-3697-1541]{N. Buchschacher}
\affiliation{\GenevaObservatory}

\author[0000-0001-5701-2529]{M. Cecconi}
\affiliation{\INAFBrenaBaja}

\author[0000-0002-9003-484X]{D. Charbonneau}
\affiliation{\CfA}

\author[0000-0003-1784-1431]{R. Cosentino}
\affiliation{\INAFBrenaBaja}

\author[0000-0003-4702-5152]{A. Ghedina}
\affiliation{\INAFBrenaBaja}

%\author{A. G. Glenday}
%\affiliation{\CfA}

\author{M. Gonzalez}
\affiliation{\INAFBrenaBaja}

\author{C-H. Li}
\affiliation{\CfA}

\author[0000-0002-5432-9659]{M. Lodi}
\affiliation{\INAFBrenaBaja}

\author[0000-0003-3204-8183]{M. L\'opez-Morales}
\affiliation{\CfA}

%\author{C. Lovis}
%\affiliation{\GenevaObservatory}

%\author{M. Mayor}
%\affiliation{\GenevaObservatory}

\author[0000-0002-9900-4751]{G. Micela}
\affiliation{\INAFPalermo}

\author[0000-0002-1742-7735]{E. Molinari}
\affiliation{\INAFBrenaBaja}
\affiliation{\INAFCagliari}

\author{F. Pepe}
\affiliation{\GenevaObservatory}

%\author{G. Piotto}
%\affiliation{\PadovaPhysicsAstronomy}
%\affiliation{\INAFPadova}

\author[0000-0003-1200-0473]{E. Poretti}
\affiliation{\INAFBrenaBaja}
\affiliation{\INAFBrera}

\author[0000-0002-6379-9185]{K. Rice}
\affiliation{\EdinburghAstronomy}
\affiliation{\EdinburghExoplanetCenter}

\author[0000-0001-7014-1771]{D. Sasselov}
\affiliation{\CfA}

%\author{D. S{\'e}gransan}
%\affiliation{\GenevaObservatory}

\author[0000-0002-7504-365X]{A. Sozzetti}
\affiliation{\INAFTorino}

%\author{A. Szentgyorgyi}
%\affiliation{\CfA}

\author[0000-0001-7576-6236]{S. Udry}
\affiliation{\GenevaObservatory}

\author[0000-0003-0311-4751]{R. L. Walsworth}
\affiliation{\CfA}
\affiliation{\MarylandPhysics}
\affiliation{\MarylandECE}
\affiliation{\MarylandQTC}

%%% DATES %%%
\received{2020 July 31}
\revised{2021 March 18}
\accepted{2021 March 23}
\published{2021 May 28}
\submitjournal{The Astronomical Journal}

%%% ABSTRACT %%%
\begin{abstract}
Radial velocity (RV) searches for Earth-mass exoplanets in the habitable zone around Sun-like stars are limited by the effects of stellar variability on the host star. In particular, suppression of convective blueshift and brightness inhomogeneities due to photospheric faculae/plage and starspots are the dominant contribution to the variability of such stellar RVs. Gaussian process (GP) regression is a powerful tool for statistically modeling these quasi-periodic variations. We investigate the limits of this technique using 800 days of RVs from the solar telescope on the High
Accuracy Radial velocity Planet Searcher for the Northern hemisphere (HARPS-N) spectrograph. These data provide a well-sampled time series of stellar RV variations. Into this data set, we inject Keplerian signals with periods between 100 and 500 days and amplitudes between 0.6 and 2.4\ms. We use GP regression to fit the resulting RVs and determine the statistical significance of recovered periods and amplitudes. We then generate synthetic RVs with the same covariance properties as the solar data to determine a lower bound on the observational baseline necessary to detect low-mass planets in Venus-like orbits around a Sun-like star. Our simulations show that discovering planets with a larger mass ($\sim$ 0.5\ms) using current-generation spectrographs and GP regression will require more than 12 yr of densely sampled RV observations. Furthermore, even with a perfect model of stellar variability, discovering a true exo-Venus ($\sim$ 0.1\ms) with current instruments would take over 15 yr. Therefore, next-generation spectrographs and better models of stellar variability are required for detection of such planets.
\end{abstract}

%%% KEYWORDS %%%
\keywords{\href{https://vocabs.ands.org.au/repository/api/lda/aas/the-unified-astronomy-thesaurus/current/resource.html?uri=http://astrothesaurus.org/uat/1930}{Gaussian Processes regression (1930)} --- \href{https://vocabs.ands.org.au/repository/api/lda/aas/the-unified-astronomy-thesaurus/current/resource.html?uri=http://astrothesaurus.org/uat/1332}{Radial velocity (1332)} --- \href{https://vocabs.ands.org.au/repository/api/lda/aas/the-unified-astronomy-thesaurus/current/resource.html?uri=http://astrothesaurus.org/uat/498}{Exoplanets
(498)} --- \href{https://vocabs.ands.org.au/repository/api/lda/aas/the-unified-astronomy-thesaurus/current/resource.html?uri=http://astrothesaurus.org/uat/1475}{Solar activity (1475)}}

%%% INTRODUCTION %%%
\section{Introduction}

State-of-the-art radial velocity (RV) searches for low-mass, long-period exoplanets are limited by signals produced by stellar magnetic variability. An Earth-like planet in orbit around a Sun-like star in its habitable zone induces a reflex RV signal on the order of 0.1\ms. However, the presence of acoustic oscillations, magnetoconvection, large-scale magnetic structures, and other stellar processes induce RV perturbations that can exceed 1\ms\ (see \cite{2016PASP128f6001F} and references therein, \cite{Cegla2019}). This stellar variability can conceal and even mimic planetary signals in RV surveys \citep{Robertson_2020}, and has resulted in many false detections (e.g., CoRoT-7d, \cite{Haywood2014}; GJ 581d and g, \cite{Robertson440}; and Alpha Centauri Bb, \cite{2016MNRAS.456L...6R}). Furthermore, these stellar processes act on timescales between minutes and months \citep{1995A&A...293...87K, 10.1093/mnras/stx1931}. For Sun-like stars, the dominant contributions to these intrinsically driven RV perturbations are from the suppression of convective blueshift and brightness inhomogeneities modulated at the rotation period \citep{2010A&A...519A..66M, dumusque2014}. The wide range of timescales and nontrivial correlations between these processes require a sophisticated statistical framework to decouple stellar activity processes from planetary signals. \cite{Hall:2018} and \cite{haywood2020unsigned} study these effects by recovering injected planets of known properties into real data.

In this work, we use state-of-the-art Gaussian process (GP) regression to account for the temporal correlations of rotationally modulated stellar activity \citep{dumusque2017, Damasso2019}. This GP regression is trained on solar data, as measured by a purpose-built solar telescope feeding the High Accuracy Radial velocity Planet Searcher for the
Northern hemisphere (HARPS-N) spectrograph \citep{Cosentino2014} operating at the Telescopio Nazionale Galileo (TNG) in the Canary Islands \citep{Dumusque:2015, Phillips2016}. In Section \ref{section:GP_framework}, we first present the solar data and the GP regression along with the resulting fit on the daily averaged solar RVs. In Section \ref{section:Results}, we introduce synthetic planets of varying semi-amplitude and orbital period in order to determine the sensitivity of the GP regression to temperate, low-mass planet searches. We conclude in Section \ref{section:SyntheticRVs} with an analysis of the baseline of RV observations and model assumptions that are necessary for a true exo-Earth detection.

%%% GPS ON SOLAR TELESCOPE DATA %%%
\section{Methods}
\label{section:GP_framework}

\subsection{Data}

We take 5 minute disk averaged exposures of the Sun using the solar telescope and the HARPS-N spectrograph at the TNG, and use a baseline of around 800 days of near-continuous, daytime, solar spectra \citep{Dumusque:2015, Phillips2016}. Neutral density filters in the HARPS-N calibration system are used to match the solar flux to the exposure time that is set to integrate over solar $p$-modes. For each exposure, the HARPS-N Data Reduction Software (DRS) \citep{Baranne_et_al_1996, Sosnowska_2012} computes the barycenter-corrected RV with 0.4\ms\ single exposure precision and the Mt. Wilson S index \citep{Wilson_1968}, a measure of stellar magnetic activity derived from chromospheric re-emission in the core of the singly ionized Ca H and K line cores \citep{Linksy_Avrett_1970}. The resulting RVs are further reduced, as described in \cite{ACC2019}, to remove the RV signatures of the solar system planets, effects of differential extinction across the solar disk in the Earth's atmosphere, and other systematic effects due to the Earth's orbit around the Sun. Additionally, we eliminate exposures contaminated by clouds or other significant atmospheric losses. We realize signal-to-noise ratios (S/R) above 300 in most exposures. After cutting exposures contaminated by clouds, we observe variations in the S/R below the 10\% level over many months with statistical RV precision varying between 0.45 and 0.6\ms. We have not observed correlations between extracted RVs and S/R. Finally, in order to mimic the sampling of a typical stellar observing schedule while preserving the exquisite S/R of the solar telescope \citep{Phillips2016}, we compute daily averaged values of each quantity including the RVs, the S index, and the corresponding mean Julian date. While realistic stellar observing schedules would not allow for this level of averaging, which integrates over variability on minute and hour timescales, we wish to assess the best-case scenario using an ultra-high S/R data set.

%%% S INDEX %%%
\subsection{Gaussian Process Kernel}

\begin{figure*}
\begin{center}
    \includegraphics[scale=1.0]{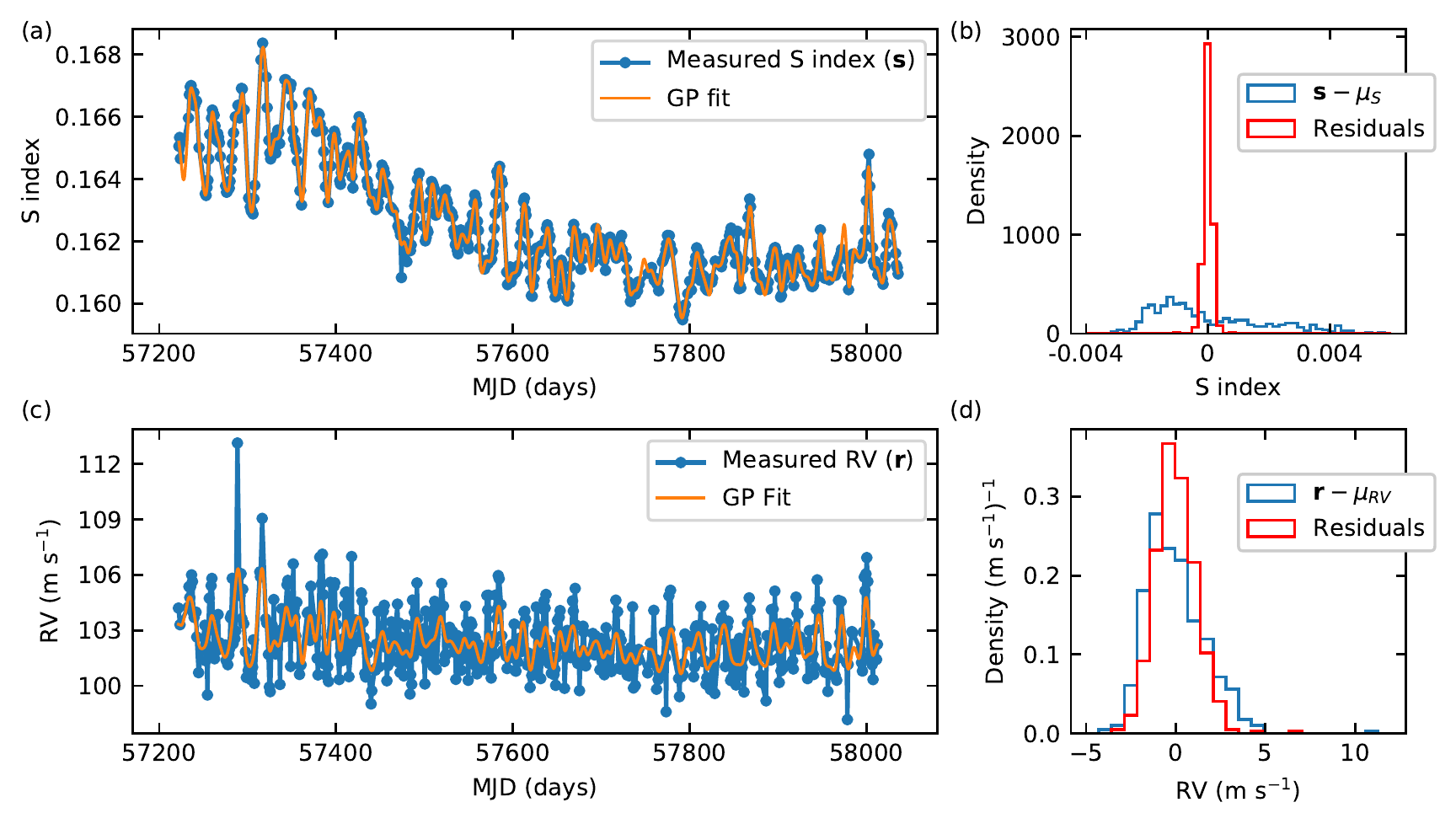}
    \caption{\label{fig:GP_fit} GP regression fits to $\sim800$ days of solar S index and RV data from the solar telescope and HARPS-N. (a) S index (blue) and GP regression fit (orange). (b) Histogram of mean-subtracted S index values (blue) and histogram of the residuals of the GP regression fit (red). (c) Daily average RVs (blue) and GP regression fit (orange). The mean value represents the HARPS-N instrumental offset and carries no physical meaning. (d) Histogram of mean-subtracted RVs (blue) and histogram of the residuals of the GP regression fit (red). The rms variation is reduced from 1.65 to 1.14\ms. The relative size of the WN compared to the correlated noise is higher in the RVs than in the S index, emphasizing the need for a more sophisticated model to account for RV variation.}
\end{center}
\end{figure*}

\begin{figure*}
\begin{center}
    \includegraphics[scale=1.0]{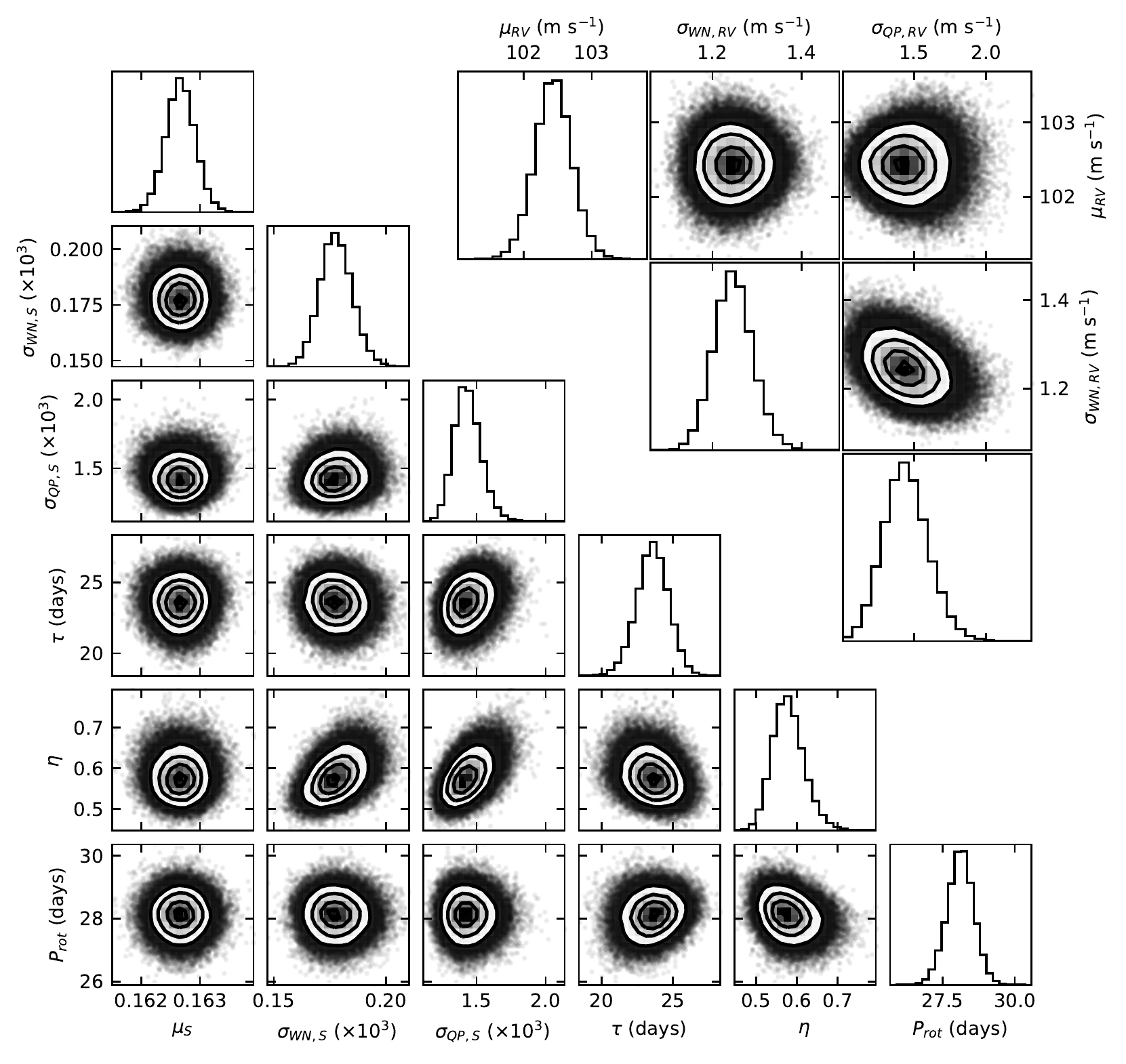}
    \caption{\label{fig:corner_plot} Marginalized distributions of the posterior distribution given by the MCMC samples. The lower left triangle shows samples (and contour lines) from the S index fit and the upper right triangle shows samples (and contour lines) from the RV fit. Each parameter is unimodal and only slight correlations exist between the parameters. Thus both posterior distributions are well behaved.}
\end{center}
\end{figure*}

Magnetic variability can be the dominant source of variance for stellar RVs of nearby, bright Sun-like and low-mass stars, reaching levels surpassing 1\ms\ \citep{Isaacson_2010, motalebi2015}. Intrinsic stellar variability introduces correlations into the RV time series that are difficult to model deterministically. GP regressions have emerged as a powerful statistical technique that relaxes the assumption of uncorrelated noise by adding nonzero terms to the off-diagonal of the data covariance matrix. Usually a kernel function is chosen to describe these covariances as a function of measurement separation time \citep{Rasmussen2006, Haywood2014, Rajpaul2015, Faria2016, Damasso2019}.

Since magnetic activity is well described by the S index and is modulated at the rotation period of the star, we fit the S index to a GP with a quasi-periodic (QP) covariance kernel function, $k_{\text{QP}}$. This kernel function is chosen heuristically to model the known properties of the magnetic activity, yielding the correlation between two measurements at times $t_i$ and $t_j$ given by
\begin{equation}
\label{eq:QP_kernel}
    k_{\text{QP}}(t_i, t_j) = \exp \left(-\frac{(t_i - t_j)^2}{2\tau^2} - \frac{1}{2\eta^2}\sin^2\left(\frac{\pi (t_i - t_j)}{P_{\text{rot}}}\right) \right),
\end{equation}
where $\tau$ is related to the average lifetime of active regions on the Sun, $\eta$ is related to the average distribution of activity in the photosphere of the Sun, and $P_{\text{rot}}$ is approximately the synodic rotation period of the Sun. \cite{Grunblatt_2015} show that this kernel function performs the best among three common GP kernel functions. The GP regression also includes a white noise (WN) term to account for additional uncorrelated noise sources such as instrumental effects. Amplitudes of both the correlated and uncorrelated terms are allowed to vary, leading to a covariance matrix, $\textbf{K}_{\text{S}}$, with components
\begin{equation}
\label{eq:K_S}
    K_{{\text{S}}_{ij}} = \sigma_{\text{QP,S}}^2 k_{\text{QP}}(t_i, t_j) + \sigma_{\text{WN,S}}^2 I_{ij},
\end{equation}
where $\sigma_{\text{QP,S}}$ and $\sigma_{\text{WN,S}}$ are the two amplitudes, and $\textbf{I}$ is the identity matrix.

\begin{table*}
\centering
\begin{tabular}{c c c c c c c}
    \hline
    \hline
    & Fit Parameter & Description & Units & Prior & MCMC Initial Guess & Fit Value \\
    \hline
    $\boldsymbol{\theta}_{\text{S}}$ (S index parameters) & $\mu_{\text{S}}$ & S index mean & & $U(0.15, 0.17)$ & $\text{mean}(\textbf{s}) = 0.1626$ & $0.1627^{+0.0002}_{-0.0002}$ \\
    & $\sigma_{\text{QP,S}}$ & correlated noise amplitude & & $U(10^{-5}, 10^{-1})$ & $\text{std}(\textbf{s}) = 1.83 \times 10^{-3}$ & $1.43^{+0.10}_{-0.09} \times 10^{-3}$ \\
    & $\sigma_{\text{WN,S}}$ & WN amplitude & & $U(10^{-5}, 10^{-1})$ & $\text{std}(\textbf{s}) = 1.83 \times 10^{-3}$ & $0.178^{+0.007}_{-0.007} \times 10^{-3}$ \\
    & $\tau$ & active region lifetime & days & $U(5, 100)$ & 22 & $23.6^{+1.1}_{-1.1}$ \\
    & $\eta$ & smoothing parameter & & $U(0.1, 0.9)$ & 0.5 & $0.58^{+0.04}_{-0.04}$ \\
    & $P_{\text{rot}}$ & solar rotation period & days & $U(24, 32)$ & 27 & $28.1^{+0.4}_{-0.5}$ \\
    \hline
    $\boldsymbol{\theta}_{\text{RV}}$ (RV parameters) & $\mu_{\text{RV}}$ & RV mean & \ms & $U(97, 113)$ & $\text{mean}(\textbf{r}) = 102.4$ & $102.4^{+0.3}_{-0.3}$ \\
    & $\sigma_{\text{QP,RV}}$ & correlated noise amplitude & \ms & $U(0.01, 10)$ & $\text{std}(\textbf{r}) = 1.65$ & $1.44^{+0.16}_{-0.15}$ \\
    & $\sigma_{\text{WN,RV}}$ & WN amplitude & \ms & $U(0.01, 10)$ & $\text{std}(\textbf{r}) = 1.65$ & $1.25^{+0.04}_{-0.04}$ \\
    \hline
\end{tabular}
\caption{Summary of parameters used for Markov chain Monte Carlo (MCMC) sampling for the S index and RV fits of solar telescope and HARPS-N data. The upper and lower bounds of the uniform priors are given in addition to the initial guesses used for each parameter. The last column shows the resulting median value of the MCMC samples and their corresponding 16\% and 84\% quantiles as error bars. \label{tbl:GP_fit_values}}
\end{table*}

\subsection{Fitting the S Index}
With a single, time-independent parameter representing the mean value of the S index, $\mu_{\text{S}}$, the set of fit parameters contains six variables:
\begin{equation}
\label{eq:S_parameters}
    \boldsymbol{\theta}_{\text{S}} = \{ \mu_{\text{S}}, \, \sigma_{\text{QP,S}}, \, \sigma_{\text{WN,S}}, \, \tau, \, \eta, \, P_{\text{rot}} \}
\end{equation}
and the GP likelihood function \citep{Rasmussen2006} is given by
\begin{equation}
\label{eq:GP_SHK_likelihood}
    \mathcal{L}\left(\boldsymbol{\theta}_{\text{S}} | \textbf{s},\textbf{t} \right) = \frac{1}{\sqrt{\det \left (2 \pi \textbf{K}_{\text{S}} \right)}} \exp \left( -\frac{1}{2} {\Delta \textbf{s}}^\text{T} \textbf{K}_{\text{S}}^{-1} \Delta \textbf{s}\right),
\end{equation}
where $\Delta \textbf{s} = \textbf{s}-\mu_{\text{S}}$ and $\textbf{s}$ is a vector containing the daily averaged S index at times ${\textbf{t}}$ for all the S index values shown in Figure \ref{fig:GP_fit}(a). Each parameter is assigned a uniform prior probability with upper and lower bounds encompassing realistic values, shown in Table \ref{tbl:GP_fit_values}. The posterior probability distribution, $p_{\text{S}}\left(\boldsymbol{\theta}_{\text{S}} | \textbf{s},\textbf{t} \right)$, is then proportional to $\mathcal{L}\left(\boldsymbol{\theta}_{\text{S}} | \textbf{s},\textbf{t} \right)$ within the prior bounds and zero otherwise.

We estimate the posterior distribution using an affine-invariant Markov chain Monte Carlo (MCMC) method, implemented with the Python \citep{Python} packages \texttt{NumPy} \citep{NumPy}, \texttt{SciPy} \citep{SciPy}, \texttt{emcee} \citep{emcee}, and \texttt{George} \citep{George}. Following \cite{emcee} 32 walkers are used to sample the parameter space, and are initialized with a normal distribution around mean values from a maximum-likelihood fit and known properties of the Sun. These values are summarized in Table \ref{tbl:GP_fit_values}. The first 100 samples are discarded allowing the walkers to converge to the posterior distribution before evaluating an additional 50,000 samples per walker. To remove correlations between samples, they are thinned by keeping one out of every $n_{\text{S}}$ samples for each walker, where $n_{\text{S}}=21$ is the average correlation length of the walkers as estimated by the autocorrelation. This yields a total of 76,032 uncorrelated samples for each parameter.

The resulting parameter estimates, $\hat{\boldsymbol{\theta}}_{\text{S}}$, are displayed in the last column of Table \ref{tbl:GP_fit_values} as the median value of these uncorrelated samples. The error bars are reported as the corresponding 16\% and 84\% quantiles. The values obtained are consistent with known solar properties. Of note is the active region lifetime, $\tau$, which is less than one rotation period despite lifetimes of photospheric faculae typically being greater than six rotation periods. Variability of the distribution of faculae on the solar surface is expected to drive this value down below the lifespan of an individual facular region. The GP regression fit to the S index and a histogram of the residuals are shown in Figures \ref{fig:GP_fit}(a), (b). The variances and covariances of the MCMC samples are shown graphically in the bottom left half of Figure \ref{fig:corner_plot}. The largest correlations are the $(\eta, \, \sigma_{\text{QP,S}})$ and $(\eta, \, \sigma_{\text{WN,S}})$ pairs with statistically significant Pearson correlation coefficients \citep{Pearson} of 0.49 and 0.42 respectively. This could indicate an unaccounted for spatial dependence in the magnetic activity. Furthermore, significant correlations exist in the $(\eta, \, \tau)$ and $(\eta, \, P_{\text{rot}})$ pairs, potentially reflecting the migration of active regions from higher latitudes (and thus longer rotation period) to lower latitudes (and thus shorter rotation period) over the 2.2 yr of data. Additionally, $\sigma_{\text{QP,S}}$ increases slightly with $\tau$, lending further evidence to a time dependence in the fit parameters as $\tau$ will be longer for the on average larger, longer-lived active regions earlier in the solar magnetic cycle. These correlations suggest a more sophisticated model may more closely represent the physical mechanisms in the photosphere of the Sun and thus capture more of the RV variation due to activity. Further investigation of these correlations are left to future study.

%%% RVS %%%
\subsection{Fitting the RVs}

We assume that the magnetic activity driving the S index also affects the observed RVs with equivalent correlated variability and thus may be described with equal GP regression parameters: $\tau$, $\eta$, and $P_{\text{rot}}$. Unlike \cite{Rajpaul2015} who fit the S index and RVs simultaneously, we model the S index and RVs consecutively in order to reduce computational complexity inherent to GPs and MCMC. Although a small improvement would be expected from simultaneous fitting, we do not anticipate this would change the overall conclusion of this work. We fit the RVs using the same QP kernel function of Equation \eqref{eq:QP_kernel} with the spot lifetime, $\tau$, spot distribution, $\eta$, and the rotation period, $P_{\text{rot}}$, fixed at the median value from the S index fit. Because the effects of solar system planets have already been removed from the RVs, we are left with only three fit parameters:
\begin{equation}
\label{eq:RV_parameters}
    \boldsymbol{\theta}_{\text{RV}} = \{ \mu_{\text{RV}}, \, \sigma_{\text{QP,RV}}, \, \sigma_{\text{WN,RV}} \}
\end{equation}
and the likelihood function becomes

\begin{equation}
\label{eq:GP_RV_likelihood}
    \mathcal{L}\left(\boldsymbol{\theta}_{\text{RV}} | \textbf{r},\textbf{t} \right) = \frac{1}{\sqrt{\det \left (2 \pi \textbf{K}_{\text{RV}} \right)}} \exp \left( -\frac{1}{2} {\Delta \textbf{r}}^\text{T} \textbf{K}_{\text{RV}}^{-1} \Delta \textbf{r}\right),
\end{equation}
where $\Delta \textbf{r} = \textbf{r}-\mu_{\text{RV}}$, $\textbf{r}$ is a vector of the RVs at times $\textbf{t}$, and $\textbf{K}_{\text{RV}}$ is populated with $\boldsymbol{\theta}_{\text{RV}}$ by substituting all `S' subscripts with `R' in Equation \eqref{eq:K_S}. Priors are shown in Table \ref{tbl:GP_fit_values}.

The posterior probability, $p_{\text{RV}}\left(\boldsymbol{\theta}_{\text{RV}} | \textbf{r},\textbf{t} \right)$, is proportional to the likelihood function within the bounds of the prior and zero otherwise. It is sampled using the same MCMC protocol as with the S index and we observe a correlation length of $n_{\text{RV}} = 11$. Retaining one of every $n_{\text{RV}}$ samples leads to a total of 145,312 uncorrelated samples. The parameter estimates, $\hat{\boldsymbol{\theta}}_{\text{RV}}$, are summarized in Table \ref{tbl:GP_fit_values} and the resulting GP regression fit and residuals are shown in Figures \ref{fig:GP_fit}(c) and (d). The fit reduces the rms scatter in the data from 1.65 to 1.14\ms, consistent with \cite{Milbourne_2019}, \cite{Miklos_2020}, and other solar analyses (see Section \ref{section:Discussion}). The marginalized distributions are shown in Figure \ref{fig:corner_plot} and again display a well-behaved posterior distribution. The only significant correlation exists between $\sigma_{\text{QP,RV}}$ and $\sigma_{\text{WN,RV}}$ and is negative, which is expected as these parameters will trade off the amount of variation seen in the RVs.

\newpage

%%% SENSITIVITY MAP %%%
\section{Sensitivity Map}
\label{section:Results}

\subsection{Synthetic Planet Model}

To explore the limits of our GP regression for detecting low-mass, long-period exoplanets, we inject synthetic planets with varying Keplerian parameters into the solar RVs. The general Doppler-induced RV, $v_{\text{RV}}(t)$, of a host star by a companion planet is given by
\begin{equation}
\label{eq:Keplerian}
    v_{\text{RV}}(t) = K_{\text{pl}} \left[ \cos (\omega + \nu (t|t_{\text{p}}, P_{\text{orb}}, e) ) + e \cos (\omega) \right],
\end{equation}
where $K_{\text{pl}}$ is the semi-amplitude, $\omega$ is the argument of periastron, $e$ is the eccentricity, $t_{\text{p}}$ is the time of pericenter passage, $P_{\text{orb}}$ is the orbital period, and $\nu(t|t_{\text{p}}, P_{\text{orb}}, e)$ is the true anomaly \citep{Perryman}. As a best-case scenario, we restrict ourselves to circular orbits (i.e., $e = 0$), which simplify the Doppler shift to the sine function
\begin{equation}
    \label{eq:simple_Keplerian}
    v_{\text{RV}}(t) = K_{\text{pl}} \sin \left(\frac{2 \pi t}{P_{\text{orb}}} + \phi \right),
\end{equation}
where $\phi$ is an arbitrary phase. A synthetic planet is then generated by choosing values for the semi-amplitude, orbital period, and phase followed by adding $v_{\text{RV}}(\textbf{t})$ to the vector of measured solar RVs, \textbf{r}, observed at times \textbf{t}. The vector the of RVs thus undergoes the transformation
\begin{equation}
    \label{eq:synthesized_RVs}
    \textbf{r} \rightarrow \textbf{r} + v_{\text{RV}}(\textbf{t}).
\end{equation}

The GP regression now contains a mean function given by the addition of the overall mean value of the RVs, $\mu_{\text{RV}}$, along with the Keplerian parameters required to describe the injected circular planetary orbit. This list of fit parameters, $\boldsymbol{\theta}_{\text{RV}}$, then
becomes
\begin{equation}
\label{eq:RV_w_planet_parameters}
    \boldsymbol{\theta}_{\text{RV}} = \{ K_{\text{pl}}, \, P_{\text{orb}}, \phi, \, \mu_{\text{RV}}, \, \sigma_{\text{QP,RV}}, \, \sigma_{\text{WN,RV}} \}
\end{equation}
and the likelihood function is still given by Equation \eqref{eq:GP_RV_likelihood}, with the exception that the vector of the fit residuals becomes $\Delta \textbf{r} = \textbf{r}- \left( \mu_{\text{RV}} + v_{\text{RV}}(\textbf{t}) \right)$ with \textbf{t} as the vector of observation times and $v_{\text{RV}}$ as the function defined in Equation \eqref{eq:simple_Keplerian}.

\subsection{Retrieval of Injected Signals}

Using the techniques of the previous section, we construct a map of detection sensitivities for a range of synthetic, low-mass, long-period planets. Our grid contains 210 injected planets with semi-amplitudes from 0.6 to 2.4\ms\ in 0.2\ms\ steps and orbital periods from 100 to 500 days in 20 days steps. The phase of each planet is drawn from a uniform distribution, $\phi \sim U(-0.1, 0.1)$ radians, and the prior on the phase is uniform from $-\pi$ to $\pi$ radians. This is done to avoid numerical instabilities associated with the phase occurring near the boundary of the prior. The prior on the semi-amplitude allows only positive values less than 10\ms\ and the prior on the orbital period is in the range $(\frac{1}{2}P_{\text{orb}}, 2 P_{\text{orb}})$. For planets detected with a high degree of statistical significance, the priors are uninformative and do not affect the results. However, for the lowest mass planets with $1 \sigma$ detections or less, the priors do constrain the results as described below. These planets set the lower bound on the range of injected semi-amplitudes used in this analysis.

\begin{figure*}
\begin{center}
    \includegraphics[scale=1.0]{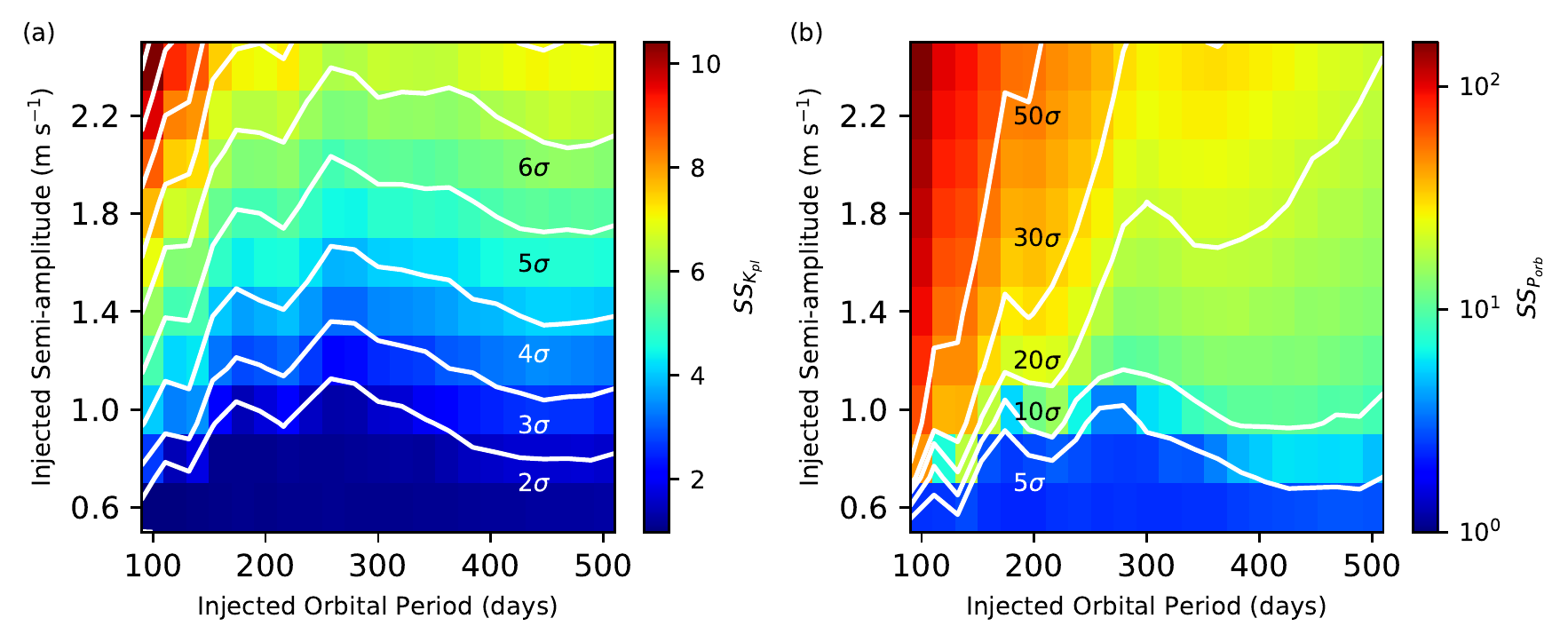}
    \caption{\label{fig:statistical_significance} Sensitivity maps of the recovered (a) semi-amplitude, $K_{\text{pl}}$, and (b) orbital period, $P_{\text{orb}}$, of synthesized planets using the GP regression with synthetic planetary signals injected into the 800 days of solar RVs. Color bars show the statistical significance (SS) of the recovered parameters as defined in Equation \eqref{eq:SS_definition}. The white lines in panel (a) show contours of $2\sigma, 3\sigma, \dots$ statistical significance and the white lines in panel (b) show contours of $5\sigma, 10\sigma, 20\sigma, 30\sigma$, and $50\sigma$ statistical significance. Each "pixel" in either image represents one of the 210 simulated planets. The orbital period is generally recovered with a high degree of confidence but the semi-amplitude is only recovered at the $5\sigma$ level for planets with semi-amplitude above 1\ms.}
\end{center}
\end{figure*}

We draw MCMC samples, exploring the semi-amplitude and orbital period linearly, in the same fashion as the previous section with the non-Keplerian parameter priors unchanged. We again define the parameter estimates, $\hat{\boldsymbol{\theta}}_{\text{RV}}$, as the median value of the uncorrelated samples and the corresponding lower and upper bounds, $\hat{\boldsymbol{\theta}}_{\text{RV, lower}}$ and $\hat{\boldsymbol{\theta}}_{\text{RV, upper}}$, as the 16\% and 84\% quantiles. The statistical significance, $SS_i$, is defined as
\begin{equation}
    \label{eq:SS_definition}
    SS_i = \frac{ \left| \hat{\theta}_{i} \right| }{\frac{1}{2} \left(\hat{\theta}_{i\text{, upper}} - \hat{\theta}_{i\text{, lower}} \right)},
\end{equation}
where $\theta_i$ is a given parameter from the vector defined in Equation \eqref{eq:RV_w_planet_parameters}. We plot the statistical significance of the recovered orbital period and semi-amplitude of the 210 synthetic planets in Figure \ref{fig:statistical_significance}. The orbital period is determined with a high degree of statistical significance, though for semi-amplitudes below 1\ms\ this should be taken as an upper bound as the MCMC samples begin to encounter the edges of the uniform prior. The semi-amplitude, however, is much less certain. For a $5 \sigma$ ``discovery'' threshold, planets with a semi-amplitude less than 1\ms\ would require more observations than the 800 days of solar telescope data used in this analysis. The structure of the contours in Figure \ref{fig:statistical_significance}(a) is likely due to our noncontinuous observing schedule. In particular, \cite{Nava_2019} find that activity-driven signals at orbital periods unrelated to either planetary orbital periods or the stellar rotation period can arise from uneven sampling. Thus even the near-daily, long-baseline observing schedule of the solar telescope decreases the semi-amplitude sensitivity \citep{Hall:2018}. We find agreement within parameter uncertainties between the injected and extracted orbital periods and semi-amplitudes, suggesting no systematic effects induced by the GP regression. In the next section we explore the baseline of observations required to detect sub-\msunit\ planets using completely synthetic RVs.

\begin{figure*}
\begin{center}
    \includegraphics[scale=1.0]{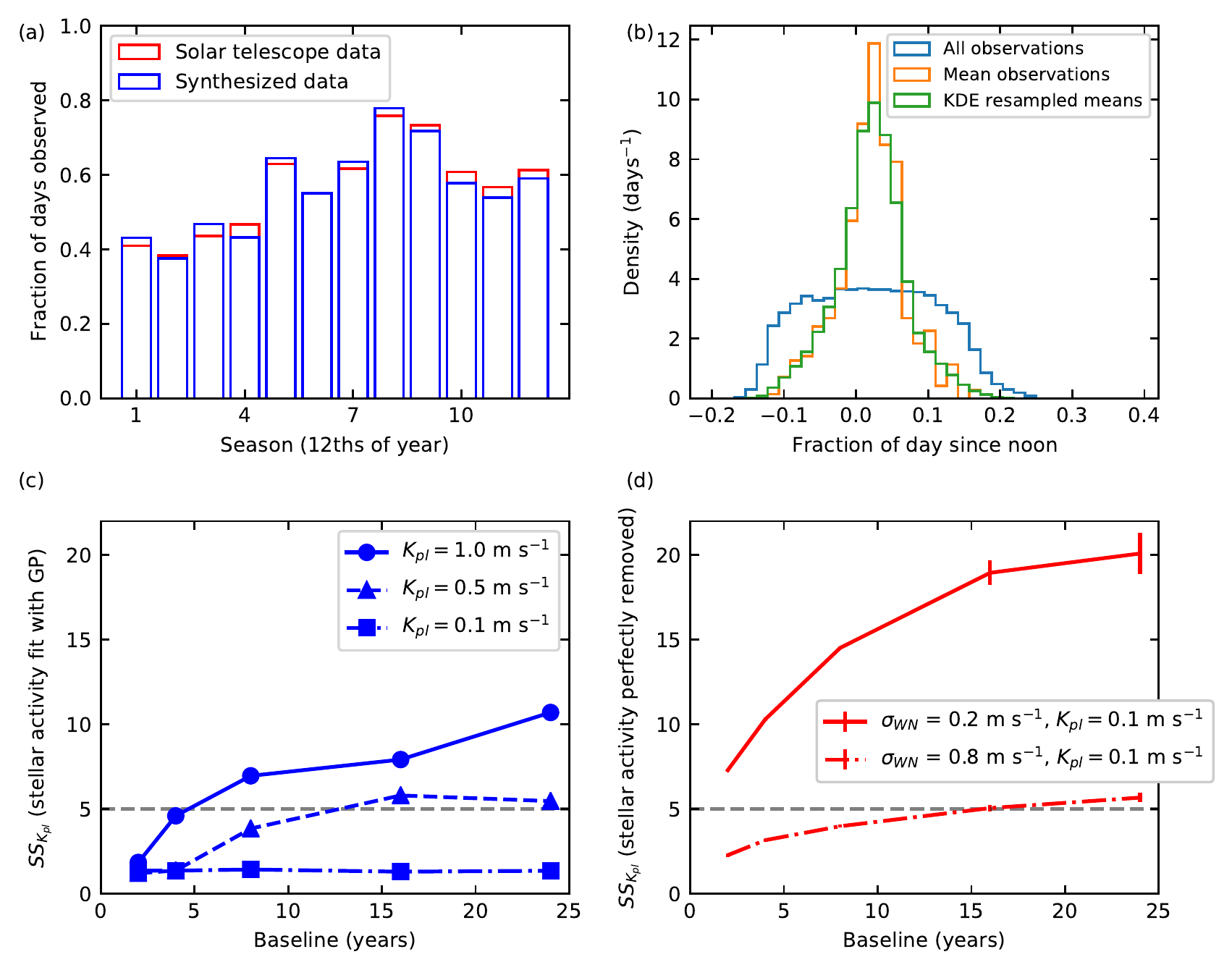}
    \caption{\label{fig:exo_venus_analysis} Generating synthetic RVs based on historical solar telescope observations. (a) Seasonal variation of solar telescope and synthetic observations. Red histogram shows the fraction of days with at least 10 5-minute exposures for 12 evenly spaced sections of a year (does not correspond to calendar months). Blue histogram shows the analogous ratio of seasonal observations from dates synthesized from the red histogram. (b) Histogram of RV observation times given as a fraction of a day since noon UTC: blue---all 5 minute solar telescope exposures, orange---daily averaged observation times for days with at least 10 5-minute exposures, and green---synthesized observation times using a Gaussian kernel density estimate. (c) Statistical significance (SS), defined in Equation \eqref{eq:SS_definition}, of the recovered semi-amplitude, $K_{\text{pl}}$, using a GP regression for an exo-Venus (225 day orbital period) as a function of baseline of data. Solid (circles) and dashed (triangles) curves show the SS for an exo-Venus with semi-amplitudes of 1.0\ms\ and 0.5\ms\ respectively. Dashed-dotted (squares) curve shows an upper bound for a semi-amplitude of 0.1\ms. Dashed gray line shows the $5\sigma$ detection threshold. (d) SS for a 0.1\ms\ exo-Venus with stellar variability perfectly removed (i.e., only WN). Solid and dashed curves show the SS for WN amplitudes of 0.2\ms\ and 0.8\ms\ respectively. Error bars are shown as vertical lines and are too small to be seen for baselines less than 15 yr. Again, the dashed gray line shows the $5 \sigma$ detection threshold.}
\end{center}
\end{figure*}

%%% SYNTHETIC RVS %%%
\section{Synthetic RVs}
\label{section:SyntheticRVs}

To study longer observing baselines than in the previous section, we synthesize not only Keplerian Doppler shifts, but also the solar RVs themselves. We thus extend the observing baseline to determine the requirements for detecting a low-mass, long-period analog. A synthetic planet with the orbital period of Venus (225 days) is injected with varying semi-amplitude. We use the orbital period of Venus to avoid systematic biases associated with measuring any periodic signal at 365 days.

%%% OBSERVING SCHEDULE %%%
\subsection{Observing Schedule}

The first step to synthesize solar telescope data is to create a realistic observing schedule to account for inclement weather and telescope downtime. We assume a naive model where each calendar month, $m$, is assigned an observation probability, $p_{\text{obs}}(m)$, which represents the fraction of days that have observations in month $m$. These probabilities are estimated using historical solar telescope data with the corresponding fraction of days with 10 or more observations. These probabilities are shown in red in Figure \ref{fig:exo_venus_analysis}(a).

For any given day $d$ in month $m$, a uniform random number, $r(d) \sim U(0, 1)$, is drawn and compared to $p_{\text{obs}}(m)$. If $r(d) \le p_{\text{obs}}(m)$, an observation occurs. This process is repeated for as many consecutive days as needed. Using this process, we create a synthetic 30 yr observing schedule mimicking the solar telescope seasonal variations. The resulting synthetic observing schedule is shown in blue in Figure \ref{fig:exo_venus_analysis}(a).

%%% OBSERVATION TIMES %%%
\subsection{Observation Times}

The second step for simulating data is to generate an observation time for each day with an observation. We begin by computing the mean observation time of all exposures in each day from historical solar telescope data, limiting only to days with at least 10 exposures. This distribution of mean observation times is histogrammed and fit to a Gaussian kernel density estimate KDE; \cite{KDE}). Synthetic observation times are drawn from this KDE. Figure \ref{fig:exo_venus_analysis}(b) shows the distribution of mean telescope exposure times and the resulting observation times drawn from the KDE. These observation times coupled with the observing days from the previous section completely determine the observing schedule of the simulated data.

%%% SYNTHESIZED RVS %%%
\subsection{Synthesized RVs}

Finally we synthesize the RVs in two steps. The first step repeats the procedure from Section \ref{section:Results} whereby a Keplerian term given by Equation \eqref{eq:simple_Keplerian} is generated with several test semi-amplitudes and a phase drawn randomly in the interval $(-0.1, 0.1)$ radians. The orbital period, however, is fixed at 225 days. The second step involves modeling the magnetic activity by using a GP with the same kernel function given in Equation \eqref{eq:K_S}. The parameter values used to populate the covariance matrix are taken from the GP regression fit of the solar data as given in Table \ref{tbl:GP_fit_values}. A random sample of RVs is then drawn from the GP using these parameter values and the synthetic observation times generated in the previous section. The sum of the Keplerian term and the GP term constitute the full RV synthesis. Note that the solar telescope RVs only cover a fraction of the solar magnetic cycle during solar minimum. As such we do not model RV variability during high activity levels. Thus, our resulting synthetic RVs represent a best-case scenario, as other Sun-like stars tend to have higher levels of magnetic activity than the Sun \citep{Reinhold_2020}. However, we emphasize that many recent studies successfully treated stellar variability by applying GP regression in RV variations of stars much more active than the Sun, such as the young pre-main-sequence stars V830 Tau \citep{Donati2016} and AU Mic \citep{Klein_2020}. In both studies, they were able to detect and characterize planetary signals with amplitudes five times smaller than the intrinsic RV variations. Therefore, we expect that an analysis including a detailed model of the solar magnetic cycle would not appreciably change the conclusions of this work, but confirmation is left to future study when precise RVs have been measured across a full solar cycle.

We then draw MCMC samples and define the parameter estimates and lower and upper bounds as before to determine the statistical significance of the recovered semi-amplitude for baselines of data equal to 2, 4, 8, 16, and 24 yr. We repeat this process for semi-amplitudes given by 1, 0.5, and 0.1 m s$^{-1}$. The resulting statistical significances are shown in Figure \ref{fig:exo_venus_analysis}(c). The dashed-dotted curve representing the 0.1\ms\ planet represents an upper bound, with the prior probability distribution on the semi-amplitude restricting its value to be positive. Even for a 0.5\ms\ planet, we determine that between 10 and 15 yr of data are needed to reach the $5\sigma$ discovery threshold using this GP regression and data similar to that of the HARPS-N solar telescope.

As shown above, a multidecade temporal baseline is required to detect a temperate, low-mass planet orbiting a Sun-like star using this GP regression. Thus we next assume that we have a direct measurement of the magnetic variability of the target star and are able to perfectly remove the effects of variability from the RVs, leaving only the planetary signals and WN. We can then fit to a sine curve with simple least-squares methods. We expect the semi-amplitude uncertainty, $\sigma_{K_{\text{pl}}}$, to scale with the measurement sensitivity, $\sigma_{\text{RV}}$, divided by the square root of the number of RV measurements, $N$ \citep{Cloutier:2018},
\begin{equation}
    \sigma_{K_{\text{pl}}} = \sigma_{\text{RV}} \sqrt{\frac{2}{N}}.
    \label{eq:wn}
\end{equation}
To confirm this scaling, we synthesize RV time series comprising simple WN and a Keplerian Doppler shift of $K_{\text{pl}}=0.1$\ms\ for an injected planet with an orbital period of 225 days (i.e., an exo-Venus). We generate this data for varying baselines and WN levels of both 0.8 and 0.2\ms. The statistical significance of the semi-amplitude, $K_{\text{pl}}$, from least-squares fits to these data sets are shown in Figure \ref{fig:exo_venus_analysis}(d) and are in good agreement with expectations set by Equation \eqref{eq:wn}. We emphasize that we need to reduce the WN to levels approaching 0.2\ms\ to reach the $5\sigma$ detection threshold in only a few years for a true exo-Venus.

%%% DISCUSSION %%%
\section{Discussion and Conclusions}
\label{section:Discussion}

The GP regression of Section \ref{section:GP_framework} reduces the rms variation of the solar RVs from 1.65 to 1.14\ms. However, treating stellar variability with this GP regression still requires 10-15 yr of densely sampled RV observations to detect long-period, low-mass planets. This is much longer than would be expected if the RVs contained purely uncorrelated WN. This result is in line with more physically motivated techniques. For example, \cite{Milbourne_2019} used magnetograms and Dopplergrams from the Helioseismic and Magnetic Imager (HMI) on board the Solar Dynamics Observatory to derive activity-driven RV time series. By modeling the HARPS-N RVs using these activity timeseries, they reduced the RV rms from 1.65 to 1.21\ms. \cite{haywood2020unsigned} reduced the RV rms to 0.85\ms\ by modeling the RVs with a linear combination of the unsigned magnetic flux from HMI and the total solar irradiance, using the $FF'$ method \citep{Aigrain2012}. Fitting our GP regression to the unsigned flux may yield a smaller variation in the RV residuals, and this will be investigated in a future work. \cite{Dumusque_2018} and \cite{Cretignier2020} used logistic regression and gradient boosting on HARPS-N solar spectra to differentiate activity sensitive and insensitive lines. Computing RVs from these sets of lines reduced the RV rms to 0.9\ms. Similarly, \cite{Miklos_2020} applied the techniques of \cite{Meunier_2017} to the HARPS-N solar RVs to estimate magnetoconvective RV variations. This analysis, however, was unable to reduce the observed rms RVs.

That the GP regression---a statistical tool---performs similarly to these more physically motivated analyses demonstrates its power: rather than requiring high-resolution solar images or specially tuned line lists, the GP regression assumes a correlation between the solar S index and observed RVs. However, the fact that the GP analysis and these physically motivated techniques arrive at the same approximately 1\ms\ RV uncertainty level indicates that something else---either another physical process operating on a different timescale or an instrumental systematic---is limiting the performance of these techniques. \cite{haywood2020unsigned} discuss some physical effects missing from the current models, to be addressed in future work. However for current data analysis models, the solar data set used in this work, with nearly 2.5 yr of near-daily observations mostly in the decline phase of Cycle 24, represents a best-case scenario; we can therefore only perform $5\sigma$ recoveries of long-period planets with semi-amplitudes greater than 1\ms, consistent with the results of the recent community-wide RV challenge of \cite{dumusque2017}.

Our work using synthetic RVs indicates that more observations will not quickly overcome this limit: as shown in Figure \ref{fig:exo_venus_analysis}, it will take 10-15 yr to reach $5 \sigma$ on a 0.5\ms\ RV signal with a 225 days period (i.e., the orbital period of Venus), and at least 25 yr for a 0.1\ms\ RV signal. The last panel of this figure also indicates that a perfect model of activity-driven correlated variations would not in itself suffice for the rapid detection of an exo-Venus or an exo-Earth; even in the absence of correlated noise, a current-generation spectrograph with a long-term stability of about 0.8\ms\ would need a 10-15 yr observing baseline to reach a $5\sigma$ detection of an exo-Earth. Successful exo-Earth discovery therefore requires both more sophisticated models of stellar variability and the improved RV precision, long-term stability, and dense observational sampling from a next-generation spectrograph \citep{NextGenSpectrographs} such as the Echelle SPectrograph for Rocky Exoplanets and Stable Spectroscopic Observations (ESPRESSO; \cite{2013PepeESPRESSO, ESPRESSO2020_2, ESPRESSO2020_1, Pepe_2021}), NEID \citep{NEID}, EXtreme PREcision Spectrometer (EXPRES; \cite{EXPRES, Blackman_2020, Brewer_2020, Petersburg_2020}), HARPS3~\citep{HARPS3}, or GMT
Consortium Large Earth Finder (G-CLEF; \cite{GCLEF}).

\newpage

%%% ACKNOWLEDGEMENTS %%%
\section{Acknowledgements}

This work was primarily supported by NASA award number NNX16AD42G and the Smithsonian Institution. The solar telescope used in these observations was built and maintained with support from the Smithsonian Astrophysical Observatory, the Harvard Origins of Life Initiative, and the TNG.

This material is also based upon work supported by NASA under grants No. NNX15AC90G and NNX17AB59G issued through the Exoplanets Research Program. The research leading to these results has received funding from the European Union Seventh Framework Programme (FP7/2007-2013) under grant Agreement No. 313014 (ETAEARTH). 

The HARPS-N project has been funded by the Prodex Program of the Swiss Space Office (SSO), the Harvard University Origins of Life Initiative (HUOLI), the Scottish Universities Physics Alliance (SUPA), the University of Geneva, the Smithsonian Astrophysical Observatory (SAO), the Italian National Astrophysical Institute (INAF), the University of St Andrews, Queen's University Belfast, and the University of Edinburgh.

This work was partially performed under contract with the California Institute of Technology (Caltech)/Jet Propulsion Laboratory (JPL) funded by NASA through the Sagan Fellowship Program executed by the NASA Exoplanet Science Institute (R.D.H.).

S.H.S. is grateful for support from NASA Heliophysics LWS grant NNX16AB79G.

A.M. acknowledges support from the senior Kavli Institute Fellowships.

A.C.C. acknowledges support from the Science and Technology Facilities Council (STFC) consolidated grant No. ST/R000824/1.

X.D. is grateful to the Branco-Weiss Fellowship for continuous support. This project has received funding from the European Research Council (ERC) under the European Unions Horizon 2020 research and innovation program (grant agreement No. 851555).

C.A.W. acknowledges support from Science and Technology Facilities Council grant ST/P000312/1.

H.M.C. acknowledges financial support from the National Centre for Competence in Research (NCCR) PlanetS, supported by the Swiss National Science Foundation (SNSF), as well as a UK Research and Innovation Future Leaders Fellowship. 

We thank Aakash Ravi for the careful preparation of figures for publication and the entire TNG staff for their continued support of the solar telescope project at HARPS-N.

%%% FACILITIES AND SOFTWARE %%%
\facilities{TNG: HARPS-N \citep{Cosentino2014}, Solar Telescope \citep{Dumusque:2015, Phillips2016}}.

\software{HARPS-N Data Reduction Software \citep{Baranne_et_al_1996, Sosnowska_2012}, Python \citep{Python}: \texttt{NumPy} \citep{NumPy}, \texttt{SciPy} \citep{SciPy}, \texttt{emcee} \citep{emcee}, \texttt{George} \citep{George}}

\bigskip{}
\bigskip{}
\bigskip{}
\bigskip{}

%%% REFERENCES %%%
\bibliographystyle{aasjournal}
\bibliography{references}

\end{document}